\documentclass[aps,prb,preprint,groupedaddress,showpacs,showkeys]{revtex4}

\usepackage{graphicx}

\begin{document}


\title{Superconducting properties of RuSr$_{2}$GdCu$_{2}$O$_{8}$ studied by SQUID magnetometry}


\author{Thomas P. Papageorgiou}
\email[Corresponding author: ]{Thomas.Papageo@uni-bayreuth.de}
\author{Hans F. Braun}
\affiliation{Physikalisches Institut, Universit\"at Bayreuth, D-95440 Bayreuth,
  Germany}
\author{Thomas Herrmannsd\"orfer}
\affiliation{Forschungszentrum Rossendorf, D-01314 Dresden, Germany}

\date{\today}

\begin{abstract}
For polycrystalline RuSr$_{2}$GdCu$_{2}$O$_{8}$ (Ru-1212), distinct peaks have been reported in d.c. magnetization 
in the superconducting state of the sample. Sr$_2$GdRuO$_6$ (Sr-2116), the precursor for the preparation 
of Ru-1212, shows similar peaks in the same temperature regime. Based on measurements performed on both bulk and powdered 
samples of Ru-1212 and Sr-2116, we exclude the possibility, that the observed behavior 
of the magnetization of Ru-1212 is due to Sr-2116 impurities. The effect is related to the superconductivity of Ru-1212, but 
it is not an intrinsic property of this compound. We provide evidence that the observation of magnetization peaks in the 
superconducting state of Ru-1212 is due to flux motion generated by the movement of the sample in 
an inhomogeneous field, during the measurement in the SQUID magnetometer. 
We propose several tests, that help to decide, whether the features observed 
in a SQUID magnetization measurement of Ru-1212 represent a property of the compound or not.     
\end{abstract}


\maketitle

\section{Introduction}

Since its first observation, high temperature superconductivity (HTSC) is a fascinating and very active field of research. 
The paper of J. G. Bednorz and K. A. M\"uller \cite{Bednorz} triggered a race for higher superconducting 
transition temperatures, \textit{T}$_{c}$, which led to the discovery of many new superconductors. All these 
compounds though, had a common feature: A layered structure, where the existence of CuO$_{2}$ layers seemed to 
be essential for the observation of superconductivity, at least until recently \cite{Maeno}. 
Because of this feature, the term \textit{Cuprates} is very often used to describe the HTSC compounds and distinguish 
them from the elemental or more conventional superconductors described by the BCS theory \cite{Bardeen}.

Although there is no complete theory explaining the superconductivity of the cuprates, the CuO$_{2}$ layers are 
believed to be responsible for conductivity and superconductivity, while interspersed layers, either insulating or weakly metallic, 
act as charge reservoirs donating carriers to the CuO$_{2}$ 
planes. Thus, the cuprates can be viewed as a stacking 
of superconducting sheets (consisting of the CuO$_{2}$  layers) coupled by Josephson interaction.

There appears to exist a relation between the maximum \textit{T}$_{c}$ of a cuprate and the number of the CuO$_{2}$ 
layers per superconducting sheet. For example \cite{books}, La$_{2-x}$M$_{x}$CuO$_{4}$ with 
M = Ba, Sr, Ca and one CuO$_{2}$ layer per superconducting sheet have a \textit{T}$_{c}$ of about 30~K, 
RBa$_{2}$Cu$_{2+m}$O$_{6+m}$ with R = Y, La, Nd, Sm, Eu, Ho, Er, Tm, Lu and two CuO$_{2}$ layers per superconducting sheet 
have a \textit{T}$_{c}$ of the order of 90~K, while Bi$_{2}$Sr$_{2}$Ca$_{n-1}$Cu$_{n}$O$_{2n+4}$ and 
Tl$_{2}$Ba$_{2}$Ca$_{n-1}$Cu$_{n}$O$_{2n+4}$ with n = 3 and three CuO$_{2}$ layers per superconducting sheet have a \textit{T}$_{c}$ 
of about 110 and 125~K respectively. Thus, the more CuO$_{2}$ layers per superconducting sheet the higher the maximum \textit{T}$_{c}$.

With this empirical relation in mind, it is interesting to investigate how the properties of a two-layer system, like 
YBa$_{2}$Cu$_{3}$O$_{6+x}$ (YBCO), would be affected, if the coupling between the superconducting sheets is changed, with the 
introduction of a metallic block. An effort to follow this idea made by L. Bauernfeind \cite{Bauernfeind1} led to the 
discovery of RuSr$_{2}$GdCu$_{2}$O$_{8}$ (Ru-1212) \cite{Bauernfeind2,Bauernfeind3}, where the (Ba,O)-(Cu,O)-(Ba,O) charge 
reservoir of YBCO is substituted by a SrRuO$_{3}$-like block. SrRuO$_{3}$ is a pseudocubic perovskite and a metallic itinerant ferromagnet 
with \textit{T}$_{Curie}$ $\sim$ 160~K \cite{Klein}.

The \textit{T}$_{c}$ of Ru-1212 depends strongly on the preparation conditions \cite{McLaughlin} and there are reports for non-superconducting samples 
\cite{Felner1} as well as for samples, in which the onset of superconductivity reaches 50~K \cite{Bauernfeind1,Bernhard1} 
or even higher for Ru/Cu substitutions \cite{Klamut2}. 
In any case, it is low compared to that of YBCO, presumably because of the underdoped character of the CuO$_{2}$ planes. 
Powder neutron diffraction studies \cite{Chmaissem,Lynn,Jorgensen} showed, that the Ru (and Gd) moments in this compound order 
antiferromagnetically at $\sim$ 135 (2.5)~K. In more detail, a canted arrangement of the moments is indicated by hysteresis loops in 
d.c. magnetization \textit{vs.} magnetic field measurements \cite{Bernhard2,Jorgensen}, which reveal a ferromagnetic component  
in the compound's magnetic properties (weak ferromagnetism). 
The fact that Ru-1212 is magnetic, is not surprising, in view of the properties of SrRuO$_{3}$ mentioned above, but makes the family of Ru-1212 
compounds \cite{Williams,Takagiwa,Hai}, 
together with RuSr$_{2}$(R$_{0.7}$Ce$_{0.3}$)$_{2}$Cu$_{2}$O$_{10}$ \cite{Bauernfeind4,Felner2,Chen,Ren}, where R = Eu, Gd (Ru-1222), 
the only family of HTSC compounds, where superconductivity arises in a state, in which magnetic order is already developed. 

There is some skepticism, especially whether superconductivity is a bulk property of Ru-1212, or even an intrinsic property of this compound at all. 
Xue \textit{et al.} \cite {Xue} report the absence of a Meissner state for Ru-1212, while Chu \textit{et al.} \cite{Chu} suggest the existence of 
a crypto-superconducting structure in this compound. On the other hand, 
heat capacity \cite{Tallon}, together with Muon Spin Rotation 
\cite{Bernhard2} and Electron Spin Resonance experiments \cite{Fainstein} indicate, that bulk superconductivity and 
magnetism in Ru-1212 coexist on a microscopic scale.

The skepticism, whether Ru-1212 is a bulk superconductor or not, is enhanced by the controversial results on the field cooled d.c. magnetization 
of Ru-1212 and related compounds published by different groups. Field expulsion shown in such a measurement, corresponding to a bulk Meissner effect, 
is generally considered as the key indicator for bulk superconductivity. Nevertheless, some published data include    
an increase of the magnetization at the temperature where 
intergranular coupling has been established \cite{Klamut2}, sometimes followed by a decrease of the magnetization at lower temperatures, which 
leads to the appearance of a peak in the magnetization \textit{vs.} temperature plots \cite{Papageorgiou}. Klamut \textit{et al.} \cite{Klamut1} 
have tentatively 
attributed these features to a change of the magnetic ordering of the Ru sublattice upon entering the superconducting state or to an anomalous 
flux lattice behavior. Artini \textit{et al.} \cite{Artini} and Bauernfeind \cite{Bauernfeind1} report a rather ``reversed'' effect, 
where a decrease of the magnetization is 
observed first, attributed to a Meissner behavior, followed by an increase of the magnetization at lower temperatures. Artini \textit{et al.} 
\cite{Artini} attribute the 
poor visibility of the Meissner state to the existence of a spontaneous vortex state proposed by Bernhard \textit{et al.} \cite{Bernhard1} or 
to a phase-lock of an aggregation of small Josephson-coupled superconducting grains or domains proposed by Chen \textit{et al.} \cite{Chen}.  
The field cooled curve of Klamut \textit{et al.} \cite{Klamut1}, with an increase of the magnetization just below \textit{T$_{c}$}, followed 
by a plateau at low temperatures is also reminiscent of a ``reversed'' effect compared to the field cooled curves published by
Bernhard \textit{et al.} \cite{Bernhard1}, which show a decrease of the magnetization just below \textit{T$_{c}$} and a plateau at low temperatures, 
considered as evidence for the existence of a bulk Meissner state in Ru-1212.

In their paper, Artini \textit{et al.} \cite{Artini} 
recognize that the complexity of the magnetic signal of Ru-1212, which consists of contributions from the Gd paramagnetic spin lattice, 
the Ru spin lattice and the diamagnetic signal due to superconductivity, can drastically affect the quality of a SQUID magnetization measurement, if 
this is done in a non-uniform field. 
Indeed, the different behaviors of the field cooled d.c. magnetization of Ru-1212 described above are reminiscent of features reported by 
McElfresh \textit{et al.} \cite{McElfresh} for an YBCO film measured in different (measured) field profiles. These features though, the specific 
characteristics of which depended on the profile of the measuring field, did not represent intrinsic properties of the sample. They did arise 
from the fact, that the magnetization of the sample was changing, because of the non-uniform field, during a measurement at fixed temperature, 
while the algorithms used by the magnetometer's software to calculate the magnetization assume that this does not happen. It is the purpose of 
this paper to investigate whether Ru-1212 shows a similar sensitivity to field inhomogeneities, which could give rise to experimental artefacts, 
during a SQUID d.c. magnetization measurement. 

\section{Experimental}\label{sec:exp}

\subsection{Sample preparation and characterization}

Polycrystalline samples of Ru-1212 were prepared following a two-step
procedure proposed by Bauernfeind \textit{et al.} \cite{Bauernfeind1,Bauernfeind3}. First, Sr-2116 was prepared from stoichiometric quantities of
RuO$_2$, Gd$_2$O$_3$ and SrCO$_3$.  The mixed powders were ground, calcined at
950~$^{\circ}$C in air, reground, milled, pressed into pellets and fired for 16~h at
1250~$^{\circ}$C in air.  In a second step, the obtained Sr-2116 was mixed
with CuO and the mixture was ground, milled, pressed into pellets and fired for 120~h at 1060~$^{\circ}$C in
flowing oxygen.

A Seifert XRD~3000~P diffractometer was used for the sample characterization. 
The powder diffraction data were recorded for 40 seconds at
each 2$\theta$ in steps of 0.01$^{\circ}$ from 5$^{\circ}$ to 75$^{\circ}$.
A weak peak indicative of SrRuO$_{3}$ trace impurities was detected in the pattern of the sample.

\subsection{Measurements}

\begin{figure}
\includegraphics[clip=true,width=75mm]{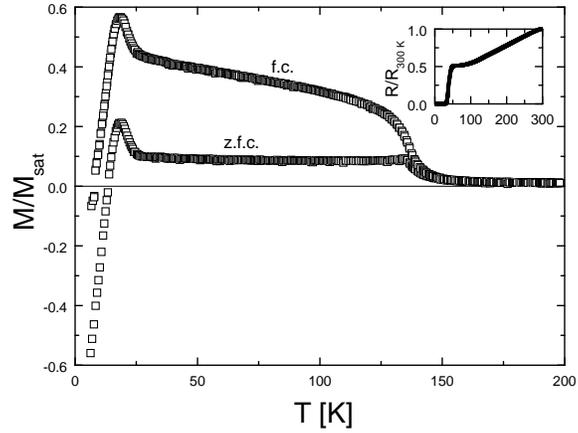}
\caption{d.c. magnetization of our Ru-1212 sample measured in a field of 0.25~mT. 
The magnetic moment M is normalised to the value of the magnetic moment M$_{sat}$ corresponding to complete flux expulsion 
from the sample. Inset: Resistance 
of the same sample normalised to the room temperature value.}\label{fig:dc}
\end{figure}

Resistance measurements were performed with a standard four-probe a.c. technique
(at 22.2~Hz) on bar-shaped pieces cut from the pellets using silver paint
contacts.  

a.c. susceptibility measurements were done with a home-made
susceptometer using a standard lock-in technique at 22.2~Hz with different field
amplitudes.

d.c. magnetization measurements were done with a commercial SQUID magnetometer (Cryogenic Consultants Ltd. S600) in the 
temperature range 7~K $\leq$ \textit{T} $\leq$ 200~K and magnetic fields \textit{B} $<$ 10~mT.
In order to overcome the problem of remanent fields, we used paramagnetic samples with high magnetic moments in low fields 
(e.g. PrCu$_{6}$) as field sensors. The magnet power supply was disconnected and an external current source 
(Knick DC-Current-Calibrator J152) was used to apply the appropriate current for the cancellation of the remanent field according to 
the signal from the paramagnetic sample. 
Complete cancelation is difficult to be achieved, but after this procedure, values for the magnetic moment of the paramagnetic samples close to the 
resolution of our SQUID (5$\cdot$10$^{-10}$~Am$^{2}$) were recorded at 7~K. 
Comparing this signal with that at the same temperature in a field of 92.9~mT we estimate a remanent field of 1.5~$\mu$T.
The low fields of figures~\ref{fig:dc} and~\ref{fig:hbeh} were also determined by a comparison of the paramagnetic sample's signal at 7~K with 
that at the same temperature in 
a field of 92.9~mT. Nevertheless, since our measurements indicate that field inhomogeneties were present, the given field values should be 
considered as estimates. For this reason, when a comparison between measurements in a certain field was necessary, the samples 
were measured one after the other, before any change of the field was undertaken.
 
\section{Results and discussion}

\begin{figure}
\includegraphics[clip=true,width=75mm]{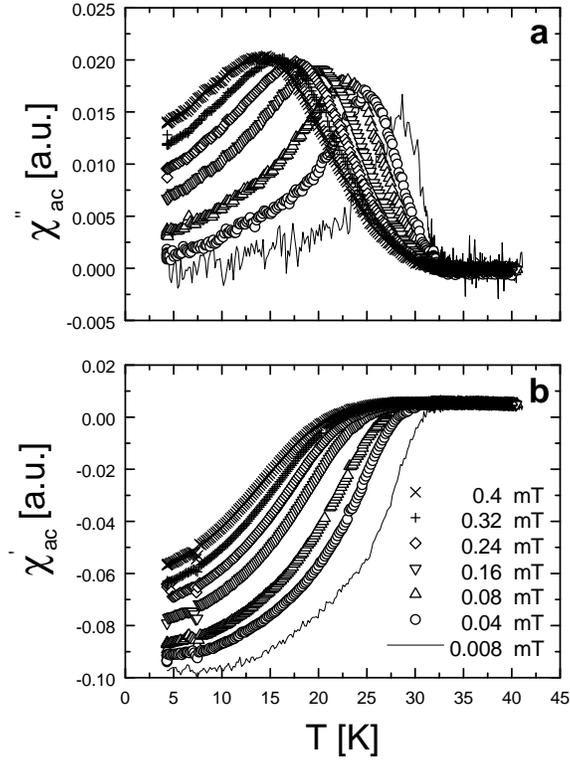}
\caption{The low temperature behavior of (a) the imaginary and (b) the real part of the a.c. susceptibility of 
our Ru-1212 sample for different field amplitudes.}\label{fig:ac}
\end{figure}

\subsection{Superconductivity and magnetism of our Ru-1212 sample}\label{SCmag}

Figure~\ref{fig:ac} and figure~\ref{fig:dc} show the a.c. susceptibility, resistivity, as well as 
the zero field cooled (z.f.c) and field cooled (f.c.) d.c. magnetization curves of our Ru-1212 sample. 
A magnetic transition is observed, as expected \cite{Chmaissem,Lynn,Jorgensen}, at 135~K. 
The hysteresis between the z.f.c. and f.c. branch of the d.c magnetization probably arises from the canting of the 
antiferromagnetically ordered Ru moments \cite{Felner1}. Increased canting of the Ru moments due to the presence of the external magnetic field 
in the f.c. process leads to higher values of the magnetization compared to the z.f.c. branch. The 
resistance of the sample, which has a metallic behavior at high temperatures, shows a cusp in the temperature range of the 
magnetic transition. The onset of 
superconductivity is at 50~K, while the resistance becomes zero at 30~K. At this temperature, the inter-granular
coupling is established and a clear diamagnetic response is observed in the
real part of the a.c. susceptibility with the corresponding loss peaks in the
imaginary part. Typical for shielding due to inter-granular coupling, the transition widens and
shifts to lower temperatures, as the a.c.  field amplitude is increased \cite{Widder}.

Distinct anomalies of the d.c. magnetization are observed for our Ru-1212 sample as it enters the 
superconducting state. The curves presented in figure~\ref{fig:dc} show 
a clear increase at 25~K and a peak at a temperature of 18~K, below which a magnetization decrease indicative of field expulsion due to superconductivity 
begins. Klamut \textit{et al.} \cite{Klamut1} were, to the best of our knowledge, first to report similar features. Their sample had a 
higher \textit{T}$_{c}$ of 35~K and the peak was observed above 25~K under the condition of zero field cooling. The origin of these peaks is still unclear. 
A comparison of figure~\ref{fig:ac} with figure~\ref{fig:dc} shows, that the onset of the peaks ($\sim$ 25~K) is close to 
the temperature, where intergranular coupling has been established. Thus, the increase of the magnetization happens when 
shielding currents start to flow between the grains. This fact was also observed by Klamut \textit{et al.} \cite{Klamut1}.  

\begin{figure}
\includegraphics[clip=true,width=75mm]{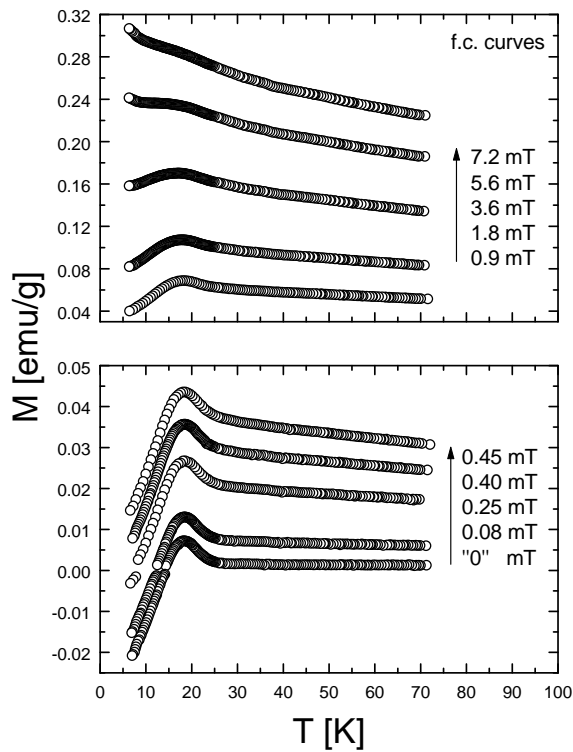}
\caption{Low temperature field-cooled d.c. magnetization of the Ru-1212 sample of figure~\ref{fig:dc} in different magnetic fields. 
The ``0'' field is the tiny remanent field, which was achived using the PrCu$_{6}$ sample as a field sensor, following the method described 
in section~\ref{sec:exp}.}\label{fig:hbeh}
\end{figure}

Figure~\ref{fig:hbeh} shows the behavior of the low temperature Ru-1212 magnetization peak in different magnetic fields. 
The position of the peak ($\sim$ 18~K) does not change significantly with magnetic field up to 3.6~mT. 
Above this field value, the feature is washed out 
and the Gd paramagnetic contribution seems to dominate at low temperatures. In these higher fields the Gd contribution 
becomes more significant also at higher temperatures, in the normal state of the sample.

In the following, we will investigate the origin of these features. We will examine first though, whether the 
observed behavior of the Ru-1212 magnetization at low temperatures is not related to impurities. 

\subsection{Can the peaks be due to Sr$_{2}$GdRuO$_{6}$ trace impurities ?}\label{sec:Sr-2116}

\begin{figure}
\includegraphics[clip=true,width=75mm]{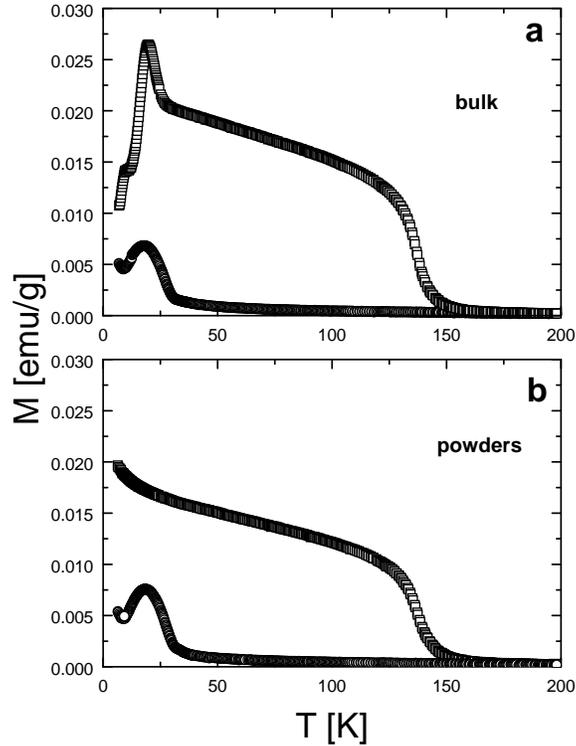}
\caption{(a) Field-cooled d.c. magnetization of bulk Ru-1212 (open squares) and Sr-2116 (open circles). The size of the peak 
for Ru-1212 in units of magnetic moment was $6.2\times 10^{-8}$~Am$^{2}$ and the mass of the sample was 10.12~mg. 
For Sr-2116 the size of the peak was $9\times 10^{-8}$~Am$^{2}$ and the mass of the sample was 14.14~mg. (b) The same after the two 
samples presented in (a) were powdered. The measuring field was 0.25~mT in all cases.}\label{fig:comp}
\end{figure}

The d.c. magnetization of Sr$_{2}$GdRuO$_{6}$ (Sr-2116) is shown in figure~\ref{fig:comp}. Sr-2116 can appear as an impurity phase in the Ru-1212 
matrix and in 
cases like ours, is used as a precursor for the preparation of Ru-1212 \cite{Bauernfeind1,Bauernfeind3,Bernhard2}. The sample was prepared as 
described in the first step of the Ru-1212 preparation. In the 
X-ray pattern traces of the original Gd$_2$O$_3$ powder were still present. The specific features of the Sr-2116 magnetization 
were presented and analysed in our previous work \cite{Papageorgiou} and will not be discussed here. It is interesting to note, that 
Sr-2116 shows a magnetization peak around 18~K, i.e., in the temperature range, where the peaks for Ru-1212 are observed.

In figure~\ref{fig:comp}a, measurements on bulk samples of Ru-1212 and Sr-2116 are shown. The results for Ru-1212 presented in figure~\ref{fig:comp}a  
and those in the previous figures are on two different pieces of Ru-1212 coming from the same pellet. All the 
necessary quantities for a comparison between the two compounds are given in the figure caption: the height of the peaks in units of magnetic moment and 
the masses of the samples. 
Although our X-ray data put an upper limit of about 0.3~mg to possible Sr-2116 impurities in our Ru-1212 sample, which represents a concentration of 
about 3 \% (resolution of the instrument), 
it is obvious, that $\sim$ 9~mg, or 88 \%, 
of Sr-2116 impurities would be needed to quantitatively explain the magnetization peak of the Ru-1212 sample as arising due to Sr-2116. 

Similar argumentation was used in our previous work 
\cite{Papageorgiou} 
to exclude the possibility that the peaks observed for Ru-1212 are due to Sr-2116 impurities. 
The fact though, that  
possible Sr-2116 grains are enclosed in a magnetic Ru-1212 matrix was not taken into account.  
We do not expect, that 
the magnetism of Ru-1212 would affect the behavior of Sr-2116 significantly
when the sample is cooled in a small applied field of 0.25~mT, like in the case presented in figure~\ref{fig:comp}. 
Assuming a homogeneous Ru-1212 matrix 
and using the magnetization M at 25~K for the bulk Ru-1212 sample from figure~\ref{fig:comp}a, we estimate a contribution to the 
magnetic field from the Ru-1212 compound 
B = $\mu_{0}$$\cdot$M = 0.2~mT, which could not enhance the magnetism of small amounts of Sr-2116, 
not visible in X-ray powder 
diffraction patterns, distributed in the Ru-1212 matrix to the level of the Ru-1212 peaks.    

We have already discussed \cite{Papageorgiou} how parameters, which can 
be affected during the preparation of Sr-2116, like oxygen content, Ru deficiencies or Cu doping (this could take place 
for example during the preparation of Ru-1212 by adding CuO to Sr-2116), could enhance the magnetism of Sr-2116 and then smaller amounts 
of this compound would be necessary 
to produce the peaks measured for Ru-1212. Indeed, the Sr-2116 peaks we found previously \cite{Papageorgiou} were more pronounced compared to  
those of figure~\ref{fig:comp}. 
These points make the above quantitative comparison somewhat uncertain. 
However, in none of the cases we studied \cite{Papageorgiou}, did we find a peak for Sr-2116-like compounds, which could explain the Ru-1212 peak, 
assuming an impurity level consistent with our X-ray data.

In order to further investigate whether the magnetization anomalies in the superconducting state of Ru-1212 are due to 
Sr-2116 impurities or not, we powdered the samples of Ru-1212 and Sr-2116 of figure~\ref{fig:comp}a and remeasured their magnetization. 
The powders were embeded in GE varnish for the 
measurements. The result is shown in figure~\ref{fig:comp}b. 
The powdered Ru-1212 sample has a  
completely different behavior at low temperatures.
The peak has dissappeared together with the magnetization decrease indicative of field expulsion 
due to superconductivity. This behavior can be attributed to grain size of the order of the penetration depth or to the quasi-two-dimensional 
character of the superconducting regions \cite{Klamut2} which prevents 
intragrain flux expulsion to occur. 
On the other hand, the properties of Sr-2116 remain unchanged after powdering. Thus, if the Ru-1212 
peak was due to Sr-2116 impurities, we expect it to still be present in the powdered sample.

A further argument is that the position of the Ru-1212 peaks is not fixed 
for all samples and seems to follow the temperature, at which intergranular coupling is established. In another case, where the Ru-1212 magnetization 
peaks have been observed \cite{Klamut1}, the sample had a higher \textit{T}$_{c}$ of 35~K compared to ours and the peak was observed above 
25~K. In this case it would be very difficult to attribute the observed peak to Sr-2116 impurities, not only quantitatively, like in our case, 
but also as far as the peak temperature is concerned.

The above analysis indicates, that the magnetization peaks observed in the superconducting state of Ru-1212 are related to the superconductivity 
of this compound. In the following we will show though, that they are not an intrinsic property of this compound, but arise from the movement of the sample in 
a non-homogeneous field during the measurement with the SQUID magnetometer. Nevertheless, we begin by discussing a model for the explanation of 
these peaks, \textit{as if they represented an intrinsic property of the compound}. This discussion will (i) help us illustrate clearly the danger of 
developing  
impressive, but invalid, explanations, when a careful check of the SQUID magnetization of Ru-1212 has not been done and 
(ii) serve as the starting point for the proposal and evaluation of several tests by which the validity of the magnetization features calculated 
by the SQUID's software for Ru-1212 (or any other superconductor) can be investigated.      

\subsection{A possible (but not real) origin of the Ru-1212 magnetization peaks}\label{PME}

Recently, the occurrence of the Paramagnetic Meissner Effect (PME) has been predicted for superconductors, when they are cooled in a 
field below \textit{T}$_{c}$, from the self-consistent solution of the Ginzburg-Landau equations 
\cite{Zharkov}. The physical picture behind this model is as follows \cite{Zharkov}: when a vortex is present inside a superconductor, the current around 
it flows in a direction to screen the vortex field from entering the bulk of the sample. In a magnetic field, additional surface current flows in 
order to prevent the field from entering the interior of the superconductor. These two currents flow in opposite directions and contribute with different 
signs to the superconductor's magnetization M. The current around the vortices gives a positive (paramagnetic) contribution, while the surface current gives 
a negative (diamagnetic) contribution. The resulting value of M can be negative or positive depending on the value of the magnetic field. The PME arises 
then, from the imbalance between the two screening currents. This physical picture can explain the magnetization peaks observed for our Ru-1212 
samples as the result of the competition between the screening currents around the vortices, which dominates at temperatures close to \textit{T}$_{c}$, 
and the surface current, the diamagnetic contribution of which starts to dominate at lower temperatures. 

\begin{figure}
\includegraphics[clip=true,width=75mm]{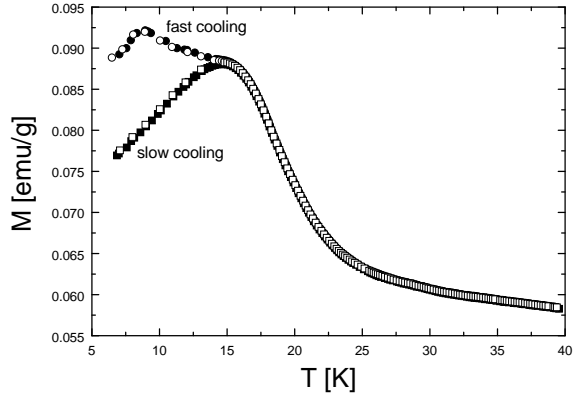}
\caption{f.c. d.c. magnetization of bulk Ru-1212 using two different cooling rates. The closed and open squares represent the two 
measurements under slow cooling conditions, while the closed and open circles the two measurements under fast cooling conditions (see text). 
For clarity several data points have been deliberately omitted below the peak position.
The measuring field was 1~mT.}\label{fig:coolrates}
\end{figure}

The model predicts the existence of vortex states with different vorticities, each of which exists in a certain range of magnetic fields. Some of these 
states are paramagnetic and some diamagnetic. Equilibrium transitions between them do not allow the observation of the PME. On the other hand, 
if metastable states exist, introduced by vortex pinning, the observation of the PME is possible. In order to investigate this feature of the model, 
we did f.c. d.c. magnetization measurements using different cooling rates.  The result is shown in figure~\ref{fig:coolrates}. 
The slow cooling measurement, which showed the peak, was 
done by cooling the sample from 200~K with a cooling rate of about 0.2~K/min below 40~K.  
The cooling rate was controlled by taking the measurements during cooldown (similar to the procedure followed for the curves of figure~\ref{fig:hbeh}) 
with a step of 0.2~K from 40 to 7~K. The curve was remeasured during warm up and no hysteresis was observed.
The fast cooling 
measurement was done by cooling the sample directly from 200~K to 7~K within 2 hours. The measurements were taken during warm up. The difference 
between the two curves is obvious below the peak position of the slow cooling measurement. The 
fast cooling measurement was repeated under the same cooling conditions. Now the measurements were not taken immediately after reaching 7~K, but after 
the sample was left at 7~K for 24 hours. This second measurement, as shown in figure~\ref{fig:coolrates}, was identical with the first one 
under fast cooling conditions. After this 
second fast cooling measurement, the slow cooling measurement was also repeated and it gave the same result as the first slow 
cooling measurement. The observed 
dependence of the measured magnetization on the cooling rate is in accordance with the model's prediction, that the observation of the PME is a signal of 
metastability. At this point we should note, that considerations similar to those developed here were recently used to introduce the PME in 
MgB$_{2}$ \cite{Horvat}. There the peak position, where the curves 
measured under different cooling rates merge, was identified as the irreversibility temperature connected with the vortex pinning 
that introduces metastability.

According to Zharkov \cite{Zharkov}, the appearance of the PME is determined by the size parameter 
A = (R/$\kappa$)(2$\pi$H/$\Phi$$_{0}$)$^{1/2}$, where R is the diameter of the sample, $\kappa$ is the Ginzburg-Landau parameter, H the magnetic field 
and $\Phi$$_{0}$ the flux quantum. For A = 1, for example, no PME is expected \cite{Zharkov}, while for A = 3 the appearance of the 
PME is allowed \cite{Zharkov}. The dependence of the PME appearance on the parameter A can explain why intergranular coupling has to be established, before the PME 
is observed. First of all, for samples with grain size of the order of the penetration depth or 
smaller, the establishment of intergranular coupling is the only possibility for the creation of the surface current, which creates the diamagnetic 
contribution to M. On the other hand, for samples with grain size bigger than the penetration depth, if this size gives a value of A not consistent 
with the appearance of the PME, then the effect will not be observed before intergranular coupling is established. So, intergranular 
coupling changes the characteristic size of the sample and can lead from A values not consistent with the 
appearance of the PME to values consistent with it. 

In the study of MgB$_{2}$, Horvat \textit{et al.} \cite{Horvat} recognize that the appearance of the PME in the z.f.c. data is difficult to understand. 
They consider this a feature, which distinguishes the PME reported for MgB$_{2}$ from the PME in the conventional or high temperature 
superconductors. 
For Ru-1212 the appearance of the PME in the z.f.c. curve 
(see z.f.c. data in figure~\ref{fig:dc}) can be understood if one keeps in mind, that Ru-1212 is a 
magnetic superconductor and even when it is cooled 
in zero magnetic field, it will react to its own magnetism. An indication for this is given by the measurement in ``0'' field presented in 
figure~\ref{fig:hbeh}, although the existence of a tiny remanent field can not be avoided. 
Since within the framework of the model presented a vortex state is necessary for the observation of the PME, 
the observation of the magnetization peak in this curve also 
can be considered indicative 
of the existence of the spontaneous vortex phase in Ru-1212, as has been already discussed for the Ruthenium-cuprates \cite{Bernhard1,Sonin}.

\subsection{The real origin of the Ru-1212 magnetization peaks}

\begin{figure}
\includegraphics[clip=true,width=75mm]{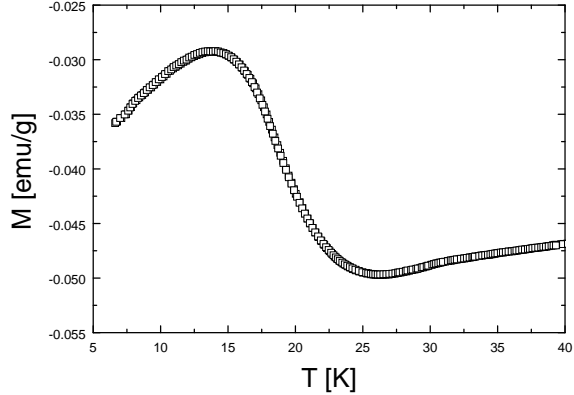}
\caption{The f.c. d.c. magnetization measurement of figure~\ref{fig:coolrates} with inversed field direction.}\label{fig:inv}
\end{figure}

In section~\ref{PME} we proposed a model for the explanation of the d.c. magnetization measurements on our Ru-1212 sample, which practically explains 
all the experimental observations. 
Within this model, the observed features in the magnetization of our Ru-1212 sample were explained as arising from the competition of a paramagnetic 
contribution from the currents around the vortices and a diamagnetic contribution from the surface current. If this is the case, we expect that 
a reversal of the field should lead to a reversal of the observed features. In figure~\ref{fig:inv}, we show the slow cooling measurement of 
figure~\ref{fig:coolrates} in a field of -1~mT. 
While the paramagnetic signal observed above 30~K is reversed as expected, 
the peak is observed again but it has not been reversed by the inversion of the field. This fact 
indicates, that the mechanism proposed in section~\ref{PME} for the explanation of the magnetization peaks observed in the superconducting state of Ru-1212 
is not the appropriate one. 

\begin{figure}
\includegraphics[clip=true,width=75mm]{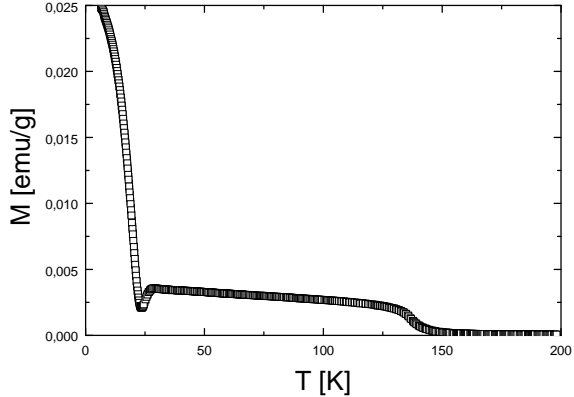}
\caption{f.c. d.c. magnetization measurement of our Ru-1212 sample in a field of 0.05~mT after the cycle 0~T $\rightarrow$ 6~T $\rightarrow$ -6~T 
$\rightarrow$ 0~T $\rightarrow$ 0.05~mT.}\label{fig:scanlength}
\end{figure}

In order to further investigate the validity of our assumptions in section~\ref{PME}, we have tested the reproducibility of our measurements 
after the superconducting magnet of the magnetometer was cycled in the following way: 0~T $\rightarrow$ 6~T $\rightarrow$ -6~T $\rightarrow$ 0~T 
$\rightarrow$ 0.05~mT. The f.c. measurement after 
this cycle is shown 
in figure~\ref{fig:scanlength}. It is obvious that the effect has reversed sign compared to figure~\ref{fig:dc}. Now a decrease of 
the magnetization is observed below \textit{T}$_{c}$ followed by a rapid increase of the magnetization at lower temperatures.

The result presented in figure~\ref{fig:inv} is similar to an observation made by Blunt \textit{et al.} \cite{Blunt}, when they investigated the 
origin of ``paramagnetic'' moments in the superconducting state of (Tl$_{0.5}$V$_{0.5} $)Sr$_{2}$(Ca$_{0.8}$Y$_{0.2}$)Cu$_{2}$O$_{y}$. On the other 
hand, the result of figure~\ref{fig:scanlength} is reminiscent of effects reported by McElfresh \textit{et al.} \cite{McElfresh} for an YBCO thin 
film, where field profiles symmetric with respect to a set value of the magnetic field created magnetization features with reversed signs 
in the superconducting state of the sample. We note, that the cycling of the magnet to high fields, before the measurement of figure~\ref{fig:scanlength}, 
is very likely to have changed the profile of the field compared to that of measurements presented in previous figures \cite{McElfresh}.

In both cases reported above, the authors \cite{Blunt,McElfresh} have interpreted their observations as artefacts arising from the movement of the sample 
in a non-homogeneous field during the measurement with the SQUID magnetometer. The problem is as follows: 
For many of the commercially available magnetometers the measurement requires the motion of the sample through a pickup coil system. 
These coils are wound in a second derivative configuration, where the two outer detection loops, located at a distance A from the center of the 
magnetometer's magnet, are wound oppositely to the two central loops located at the center of the magnet. During the measurement, 
the movement of the sample through the pickup coils induces currents in the detection loops, which, through an inductance L, create magnetic flux 
in the SQUID circuit, resulting in an output voltage V, which depends on the position of the sample z. This signal V(z) is fitted by the SQUID's 
software for the determination of the sample's magnetic moment. Nearly all analysis methods of the V(z) signal make two significant assumptions for 
the magnetic moment of the sample: (a) that it is approximated by a magnetic dipole moment and (b) that the sign and value of this moment do not change 
during the measurement. A superconducting sample though, will follow a minor hysteresis loop during the measurement, when the magnetometer's field 
is not homogeneous. Libbrecht \textit{et al.} 
\cite{Libbrecht}, using the Bean model \cite{Bean},have shown that, 
depending on the 
sample properties and the level of field inhomogeneity, it is even possible, that the sample's magnetization will change sign and eventually reach its 
original value with reversed sign at the end of the scan in a non-homogeneous field. Such problems creating spurious signals in the d.c. magnetization of 
superconducting materials have been discussed in the past by several authors \cite{McElfresh,Blunt,Libbrecht,Schilling,Braunisch}. 

\begin{figure}
\includegraphics[clip=true,width=75mm]{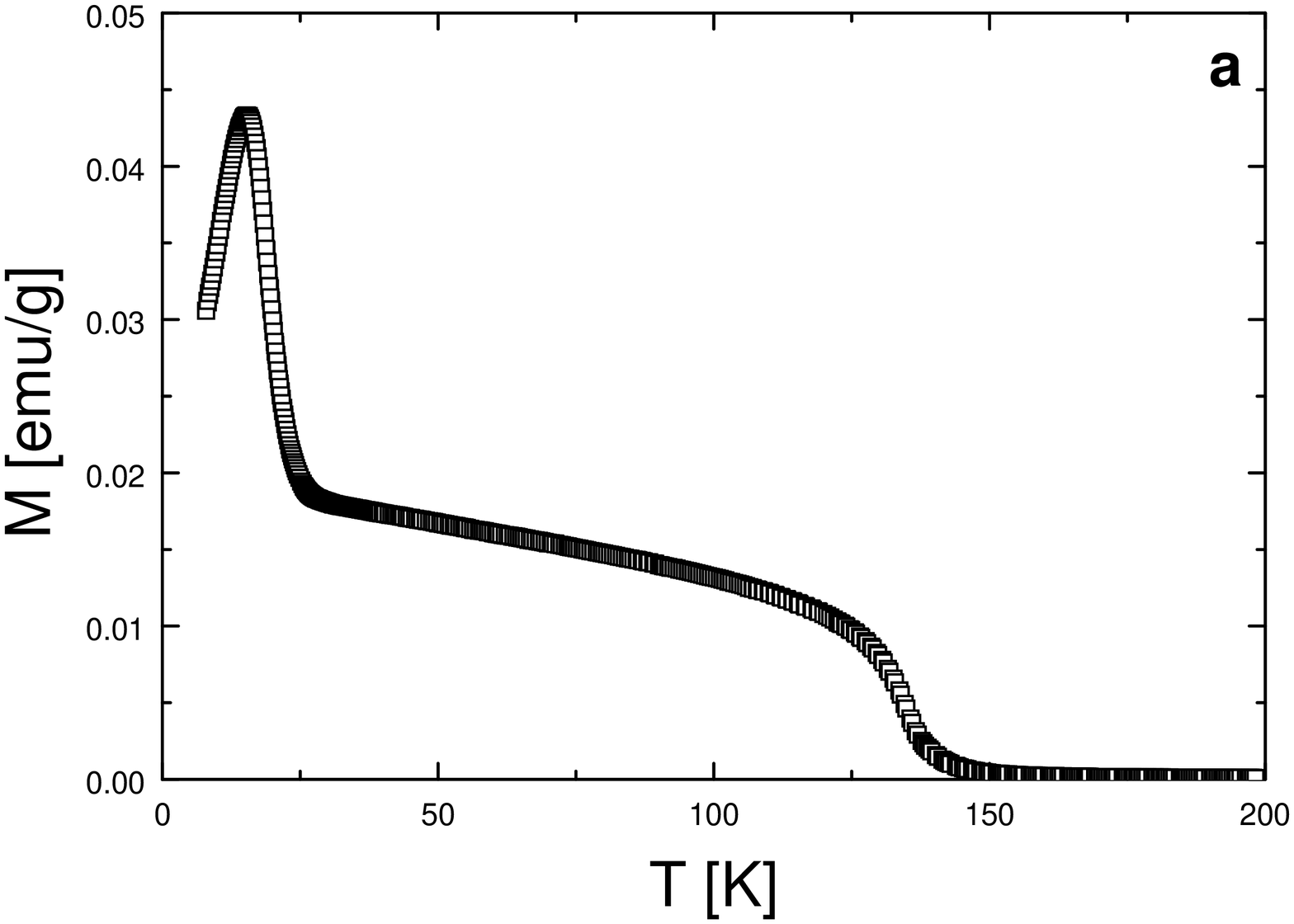}
\includegraphics[clip=true,width=75mm]{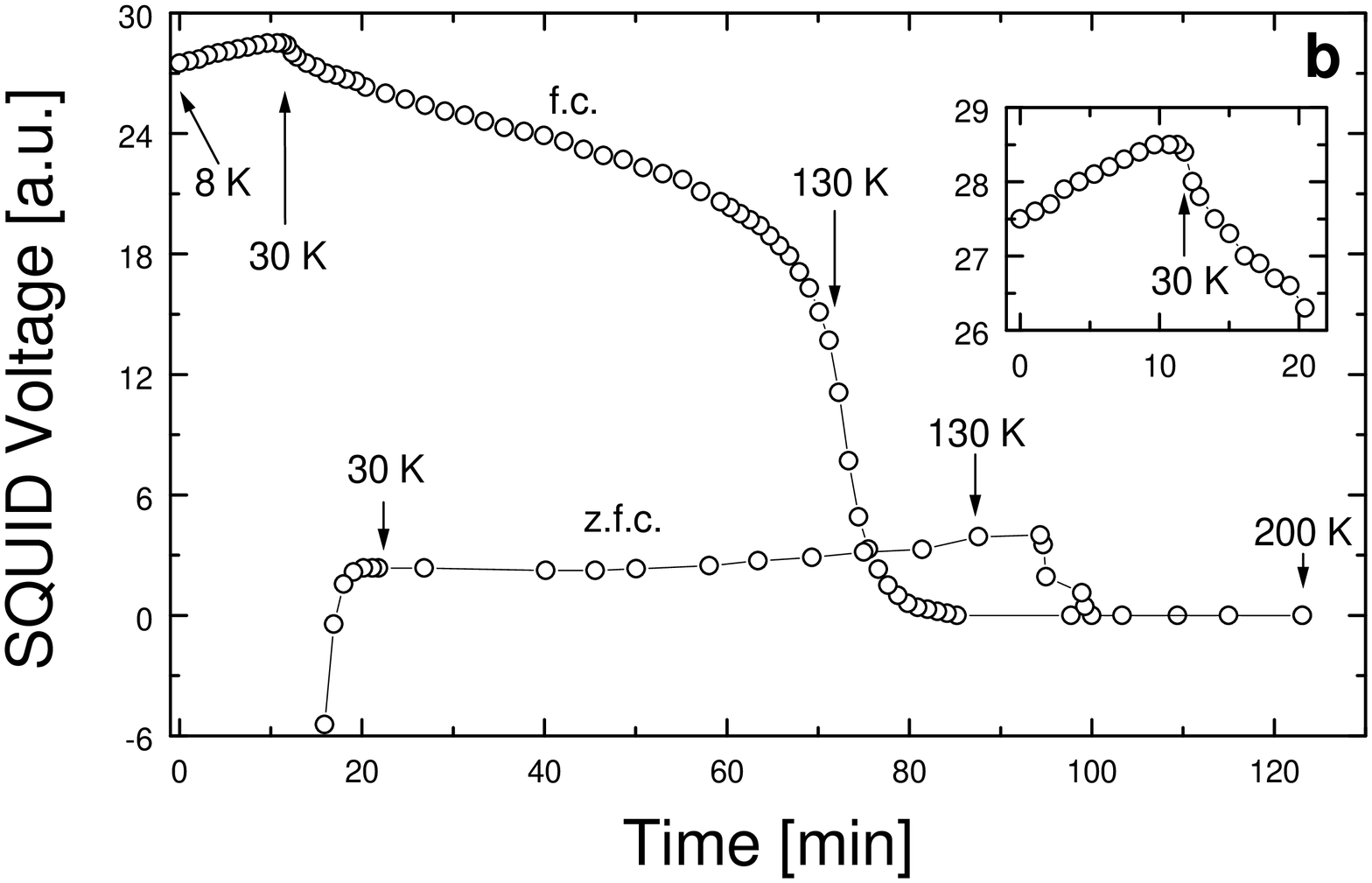}
\caption{(a) f.c. d.c. magnetization measurement of the moving Ru-1212 sample, where the peak like feature at low temperatures is obvious. The 
measurement was done in a nominal field of 0.25~mT. (b) The output voltage of the SQUID circuit with stationary sample in the same field. 
Characteristic temperatures given
at certain time points are derived from a thermometer located in the flowing He gas a few centimeters away from the sample and should be 
considered as a guide to the eye. The data are corrected for the voltage \textit{vs.} time drift of the SQUID setup. Inset: the low temperature 
part of the f.c. measurement.}\label{fig:SQUIDV}
\end{figure}

The above analysis motivated us to investigate whether Ru-1212, in its superconducting state, shows a similar sensitivity to field inhomogenieties, 
which could create artefacts in its measured magnetization. We worked in the following way: another cycle of the magnet from high magnetic fields to zero 
and finally to 0.25 mT was done. After this cycle the peak in the 
f.c. d.c. magnetization data reappeared, as shown in figure~\ref{fig:SQUIDV}a. After this measurement and without changing the field, we placed the sample 
in the center of the pickup coil system and recorded the output voltage of the SQUID circuit as the sample was cooled from 200~K without moving it. 
With this type of measurements absolute values of the magnetization can not be calculated, however, the recorded signal is 
proportional to the magnetization of the sample. The result is shown in 
figure~\ref{fig:SQUIDV}b. No peak like feature is observed at low temperatures in this measurement. Only a decrease of the magnetization 
indicative of field expulsion 
as the sample is entering the superconducting state is observed. In the same figure we have also included the z.f.c. 
measurement for our sample. Again, no peak like feature of the magnetization at low temperatures is observed. The measurements presented in 
figure~\ref{fig:SQUIDV} clearly illustrate that the peak like features of the d.c. magnetization observed in the superconducting state of 
Ru-1212 are an experimental artefact arising from the movement of the sample in an inhomogeneous field during the measurement in the SQUID 
magnetometer. 

In view of the observed sensitivity of the Ru-1212 measured magnetization to field inhomogenieties and
since we have no evidence that a field reversal, similar to that for the measurements presented in figure~\ref{fig:inv}, or cycling of the magnet, 
similar to that before the measurements presented in figure~\ref{fig:scanlength}, affect the field profile of the superconducting magnet in a systematic way,  
we consider measurements similar to those in figure~\ref{fig:SQUIDV}b as the most reliable test, that helps to decide, whether the observed features 
in a SQUID magnetization measurement of Ru-1212 (or any other superconductor) represent a true property of the compound or not. 

\subsection{Some remarks for Sr-2116}

\begin{figure}
\includegraphics[clip=true,width=75mm]{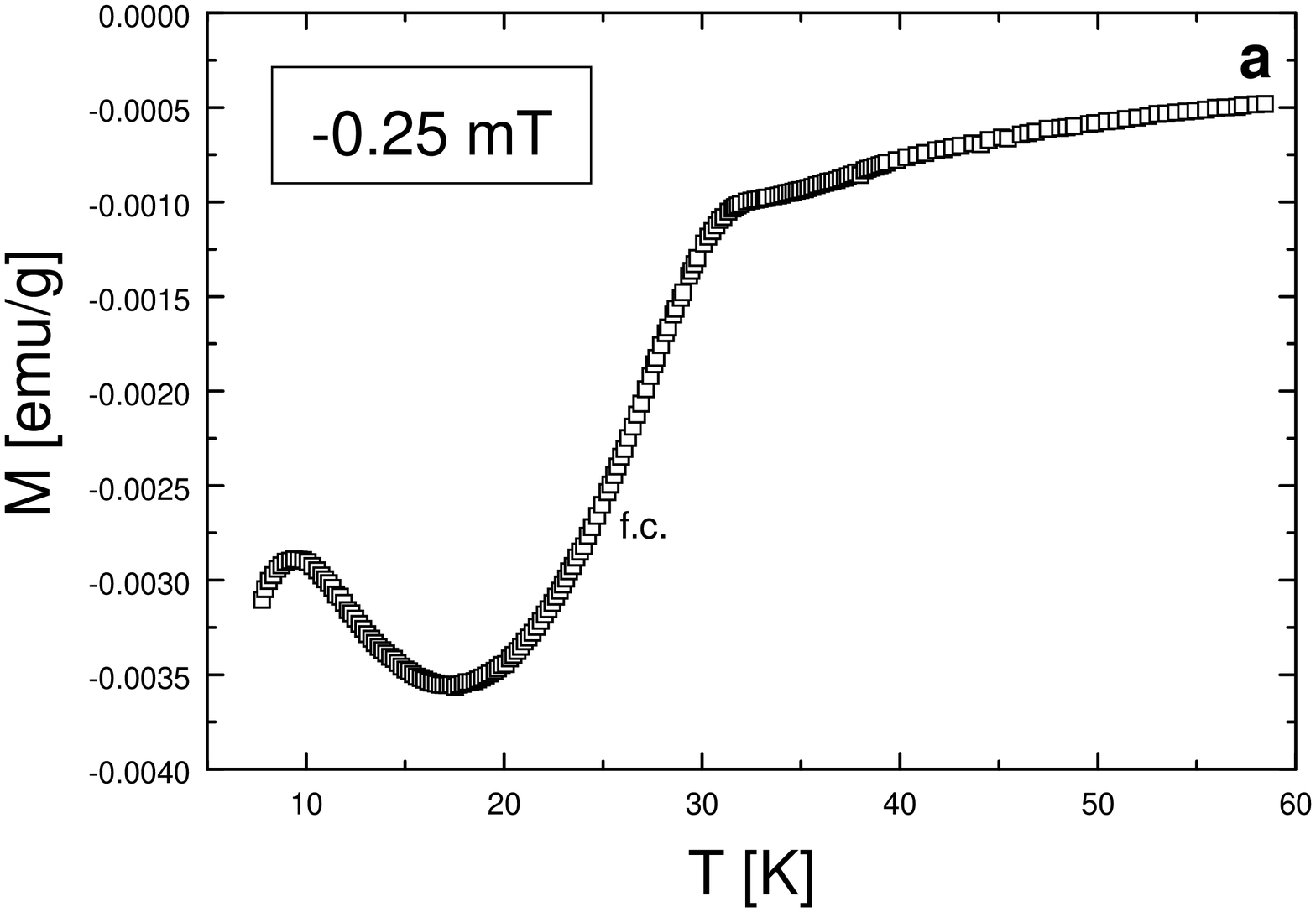}
\includegraphics[clip=true,width=75mm]{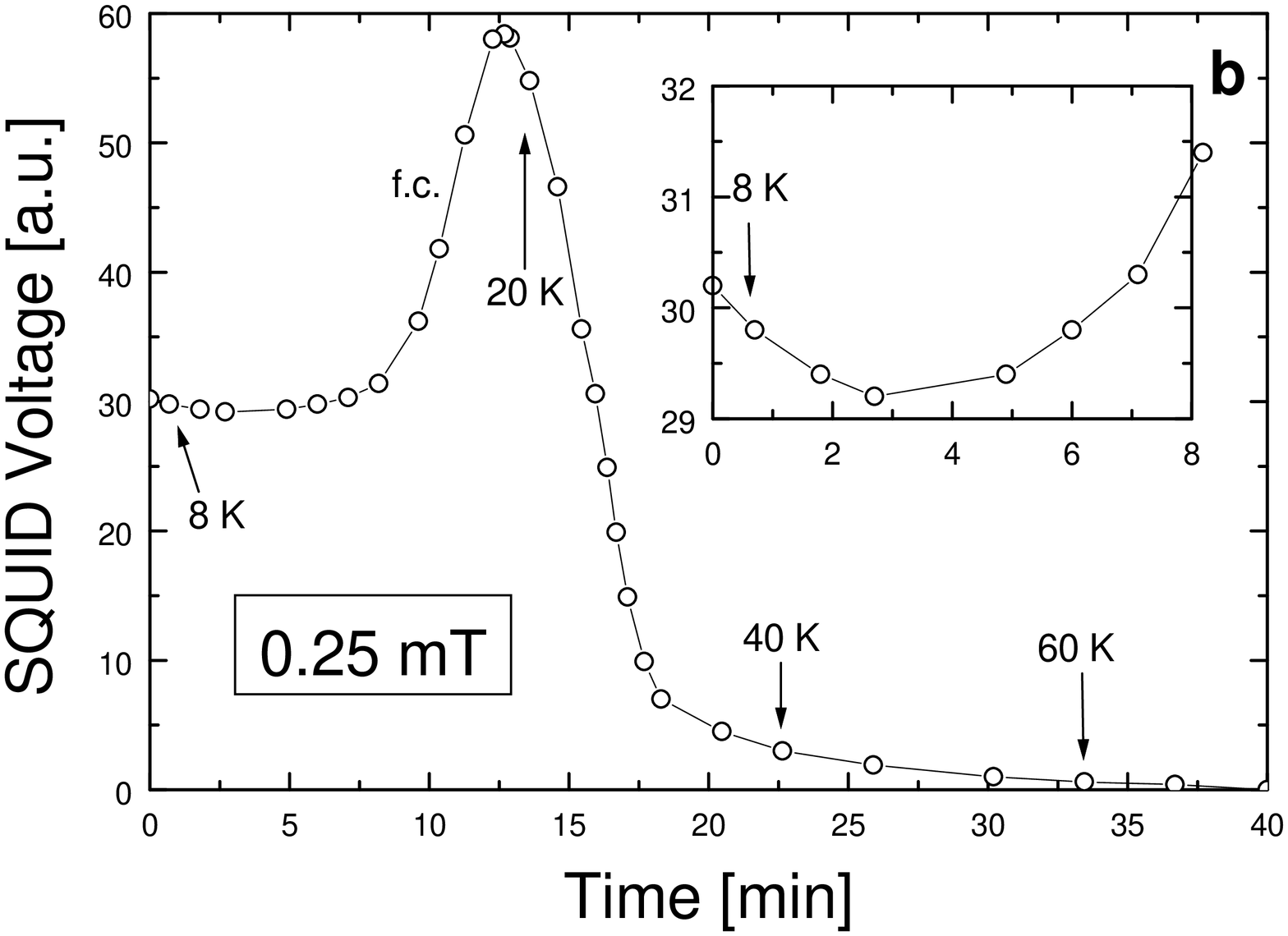}
\caption{(a) f.c. d.c. magnetization measurement of the moving Sr-2116 sample in a field of -0.25~mT. (b) The output voltage of the SQUID circuit 
with stationary sample in a field of 0.25~mT. Characteristic temperatures given at 
certain time points are derived from a thermometer located in the flowing He gas a few centimeters away from the sample and should be 
considered as a guide to the eye. The data are corrected for the voltage \textit{vs.} time drift of the SQUID setup. Inset: the low temperature 
part of the f.c. measurement.}\label{fig:Sr-2116V}
\end{figure}
 
The field profile of the SQUID's superconducting magnet can affect the measured magnetization not only of superconducting samples (like Ru-1212) but 
also of any sample showing hysteresis in its magnetization.  
This sets the validity of the reported 
magnetization peak for Sr-2116 under question (see section~\ref{sec:Sr-2116} and reference [28]). Thus, we tested Sr-2116 as introduced above. 
In figure~\ref{fig:Sr-2116V}a we show a f.c. measurement of a Sr-2116 sample in a negative field of -0.25~mT and in figure~\ref{fig:Sr-2116V}b a 
f.c. measurement of the SQUID setup's output voltage in a field of 0.25~mT. The peak, which has been attributed to a response of the 
paramagnetic Gd moments to the antiferromagnetic ordering of the Ru moments at about 35~K \cite{Papageorgiou}, is reversed in the negative field and 
is also obvious in in the measurement without moving the sample. These measurements show, that the reported 
\cite{Papageorgiou} magnetization peak for Sr-2116 represents a true property of the material.

\section{Concluding remarks} 

In summary, we have investigated the origin of ``paramagnetic''-like features (see figures~\ref{fig:dc} and ~\ref{fig:scanlength}) in the superconducting 
state of Ru-1212. We have shown, that these features, with high probability, are experimental artefacts arising from the movement of the sample 
in an inhomogeneous field during 
the measurement with the SQUID magnetometer. In view of the observed sensitivity 
of the Ru-1212 measured properties to the field profile of the superconducting magnet, we have 
proposed and evaluated several tests, which can help to decide whether the observed features in a SQUID magnetization measurement of Ru-1212 is an 
intrinsic property of the compound or not, with the most reliable one being the recording of the SQUID circuit's output 
voltage as the sample is cooled or warmed up in the magnetometer without being moved. We consider these tests as general tests, which can be used 
for the evaluation of the data on any superconducting sample.

Our work shows, that any ``paramagnetic''-like features in the 
superconducting state of Ru-1212, which could easily be related to some response of the Gd moments in this compound and considered as an indication 
for the lack of a Meissner state for Ru-1212, have to be evaluated 
carefully. On the other hand, features indicative of the existence of the Meissner state for this compound have also to be investigated carefully. 
Our work together with the measurements of McElfresh \textit{et al.} \cite{McElfresh} indicate, that any measured ``paramagnetic''-like features can 
easily be turned to ``diamagnetic''-like features, which could be mistaken as 
an indication of the existence of a bulk Meissner state for Ru-1212, by a reversal of the field profile with respect to the set value of 
the magnetic field. The f.c. measurement on a stationary sample
presented in figure~\ref{fig:SQUIDV}b indicates, that the Meissner state indeed exists for Ru-1212. Nevertheless, similar measurements, where the 
calculation of the absolute values of the magnetization will be possible for an estimation of the superconducting volume of the sample, are necessary, 
in order to clarify this issue.     

Both here and in our previous work \cite{Papageorgiou} we have shown, that Sr-2116 can also be a good candidate for the explanation 
of any ``peculiar'' features in the low temperature properties of Ru-1212 like d.c. magnetization and specific heat. Since new Ru-1212 
compounds with other lanthanides in the place of Gd have recently been reported \cite{Williams,Takagiwa,Hai,Ruiz}, 
we note, that there is a whole series of Sr-2116 compounds also with other lanthanides in the place of Gd, 
which are magnetic at low temperatures \cite{Battle1,Battle2,Battle3,Battle4,Battle5,Doi}. Some of these Sr-2116 
compounds doped with Cu, like Sr$_2$YRu$_{1-x}$Cu$_{x}$O$_6$ (but not Sr$_2$GdRu$_{1-x}$Cu$_{x}$O$_6$ \cite{Papageorgiou}) are also reported 
superconducting with a \textit{Tc} similar to those of the Ru-1212 compounds \cite{Wu1,Chen2,Wu2,Blackstead,Wu3}. Careful studies, similar to 
those in our previous work \cite{Papageorgiou} and in section~\ref{sec:Sr-2116} are necessary to exclude the possibility, that trace impurities 
of Sr-2116 compounds are responsible for the properties attributed to the Ru-1212 phase. 

\begin{acknowledgments}

We would like to thank Dr. Alexander Schindler for useful discussions.

\end{acknowledgments}

\end{document}